\documentclass[sigconf, screen, nonacm]{acmart}

\AtBeginDocument{%
  }

\acmConference[LOCO'24]{1st International Workshop on Low Carbon Computing}{December 3, 2024}{Glasgow, United Kingdom}
\acmISBN{}
\acmDOI{}
\setcopyright{none}

\usepackage{amsmath}
\usepackage{mathtools}
\usepackage{hyperref}
\usepackage{todonotes}

\newcommand{\parag}[1]{\vspace{1mm plus 1mm}\noindent\textbf{#1.}\hspace{1mm}}

\begin{document}

\title{Choosing the Right Battery Model for Data~Center~Simulations}

\author{Paul Kilian}
\orcid{0009-0003-1474-1723}
\affiliation{%
  \institution{KTH Royal Institute of Technology}
  \city{Stockholm}
  \country{Sweden}
}

\author{Philipp Wiesner}
\orcid{0000-0001-5352-7525}
\affiliation{%
  \institution{Technische Universität Berlin}
  \city{Berlin}
  \country{Germany}
}

\author{Odej Kao}
\orcid{0000-0001-6454-6799}
\affiliation{%
  \institution{Technische Universität Berlin}
  \city{Berlin}
  \country{Germany}
}

\begin{abstract}
    As demand for computing resources continues to rise, the increasing cost of electricity and anticipated regulations on carbon emissions are prompting changes in data center power systems. 
    Many providers are now operating compute nodes in microgrids, close to renewable power generators and energy storage, to maintain full control over the cost and origin of consumed electricity.
    Recently, new co-simulation testbeds have emerged that integrate domain-specific simulators to support research, development, and testing of such systems in a controlled environment.
    Yet, choosing an appropriate battery model for data center simulations remains challenging, as it requires balancing simulation speed, realism, and ease of configuration.

    In this paper, we implement four different battery models for data center scenarios within the co-simulation framework Vessim and analyze their behavior.
    The results show that linear models, which consider inefficiencies and power limits, closely match the behavior of complex physics-based models in short-term experiments while offering faster execution, and not requiring knowledge on electrochemical reactions and circuit-level dynamics. 
    In contrast, simple, lossless models fail to accurately represent complex behavior and provide no further runtime advantage.
\end{abstract}

  
\maketitle

\let\thefootnote\relax\footnotetext{Presented at the 1st International Workshop on Low Carbon Computing (LOCO) in Glasgow, Scotland, UK, on 3 December 2024}

\section{Introduction}
As the adoption of modern data-intensive technologies such as artificial intelligence continues to accelerate across industries, the need for computational power is rising steadily~\cite{Wu_SustainableAIMLSys_2022, Payal_NatureCarbonImpactAI_2020}, leading to increased power usage of today's data centers~\cite{IEA_DCEnergy_2022, masanet2020recalibrating}. Recent work shows that improvements in hardware and software efficiency is unlikely to alleviate future power demands~\cite{IEA_DCEnergy_2022, Bashir_CaseForVirtualizingEnergySystem_2021, masanet2020recalibrating}, which puts pressure on operators to find other ways to decrease their carbon footprint with potential carbon pricing mechanisms on the horizon~\cite{boyce2018carbon, best2020carbon}.
These developments have shifted the focus towards a new paradigm called \textit{carbon-aware computing} in both academia~\cite{Radovanovic_Google_2022, Anderson_Treehouse_2022, Hanafy_CarbonScaler_2023} and industry~\cite{Google_CarbonAware, XBox_CarbonAware, VMWare_CarbonAware, Windows_CarbonAware}, aiming to align the power usage of computing resources with local power production of  renewable energy sources. This is made possible because many computing workloads are inherently flexible in both location and time of execution, allowing them to be distributed across datacenters and scheduled for execution at variable times.

However, despite a growing number of data centers being operated in microgrid environments~\cite{google2024microgrid, microsoft2024microgrid, migrogridknowledge2024}, frequently comprising energy storage, most approaches in carbon-aware computing do not consider any form of local energy storage systems~\cite{Wiesner_QualityAdaptation_2025, Radovanovic_Google_2022, Zheng_MitigatingCarbonLoadMigration_2020, wiesner2024fedzero, Hanafy_GoingGreenLessGreen_2024, Hanafy_CarbonScaler_2023, gsteiger2024caribou, murillo2024cdn_shifter, souza2023casper}. A few notable exceptions include the CarbonExplorer framework~\cite{Acun_CarbonExplorer_2023} or the DataZero project~\cite{SEPIA_DATAZERO_2019}.
This is partly because existing simulation testbeds have very limited capabilities for modeling battery systems, restricting evaluation opportunities. 
Moreover, little research has been conducted on selecting appropriate battery models for different simulation use cases, despite the significant trade-off between simple, efficient models that lack physical accuracy and physics-based models that are complex to parameterize and computationally expensive.

In this work, we aim to define interfaces for integrating battery models of varying complexity into a microgrid co-simulation and analyze their performance. Our contributions are:
\begin{itemize}
    \item we extend Vessim~\cite{Wiesner2024_vessim}, a co-simulation framework for energy-aware and carbon-aware computing systems, to enable the step-wise execution of external battery simulators.
    \item we implement three new battery models within Vessim, a state-of-the-art linear model of battery behavior~\cite{kazhamiaka2019tractable} and two physics-based models~\cite{Sulzer_PyBaMM_2021, tranter2022liionpack}.
    \item we analyze the behavior and runtime of the existing and newly implemented models in simulated scenarios
\end{itemize}

The remainder of this work is structured as follows.
Section~\ref{sec:rw} reviews related work. 
Section~\ref{sec:battery_simulator_integration} extends Vessim with a standardized battery interface.
Section~\ref{sec:models} details the integrated battery models, ranging from simple linear to physics-based approaches.
Section~\ref{sec:evaluation} evaluates the battery models, analyzing accuracy, efficiency, and grid impact. 
Section~\ref{sec:discussion} discusses our findings.
Section~\ref{sec:outlook} concludes the paper with future research directions.

\section{Related Work}
\label{sec:rw}

We survey related works on energy management approaches in microgrids, energy system simulation in the context of carbon-aware computing, and physics-based battery modeling.

\parag{Microgrid energy management}
Although there exists some research on studying energy management of microgrid systems~\cite{SustainableEdgeComputing, alzahrani2017modeling, MicrogridCosim, abrishambaf2017implementation}, either they heavily rely on hardware-based modeling for their case-study simulations, or they do not discuss and explain the usage of their battery models, if present. On top of that, current approaches lack flexibility to switch individual battery models, making it difficult to examine and compare them effectively.

The DATAZERO project~\cite{SEPIA_DATAZERO_2019} focuses on design and operation of data centers, explicitly considering batteries as part of their optimization approach regarding electrical management, next to other forms of energy storage. In their small-scale experiments, however, they also relied on physical batteries using \textit{Power Hardware-In-the-Loop} instead of investigating mathematical battery models.

\parag{Energy system simulators}
Acun et al. proposed CarbonExplorer~\cite{Acun_CarbonExplorer_2023} for analyzing a data center's carbon footprint, targeting the operational as well as the embodied carbon emissions of the data center ecosystem in a holistic approach. Carbon explorer utilizes a modular C/L/C battery model~\cite{kazhamiaka2019tractable} that includes several characteristics of lithium-ion batteries to simulate their energy storage. However, the authors do not explain the reasons for utilizing this particular model for simulation purposes and also do not consider more sophisticated battery modeling.

Vessim~\cite{Wiesner2024_vessim}, on the other hand, aims to serve as a co-simulation testbed for carbon-aware applications and systems. Vessim combines domain-specific simulators while empowering users to flexibly configure their (partially) simulated power systems. It enables seamless integration with actual software and hardware, offering a range of utility functions such as replaying power production forecasts or accessing live data through queries. So far, however, the efforts regarding battery modeling have not exceeded way too simple battery models.

\parag{Physics-based battery modeling}
With the goal to support research and development in energy storage systems, Sulzer et al. created PyBaMM~\cite{Sulzer_PyBaMM_2021} (Python Battery Mathematical Modeling), as an open-source Python library for modeling and simulation of single-cell batteries. The framework offers a wide range of mathematical models to describe the electrochemical processes inside batteries, including the comprehensive physics-based Doyle-Fuller-Newman model (DFN) and the more efficient Single Particle Models (SPMs) model for lithium-ion batteries. In addition to that, PyBaMM allows users to define complex battery geometries and generates computational meshes automatically, and facilitates the parameterization of battery models by providing tools to calibrate and validate against experimental data, which enables accurate simulation of battery behavior under various operating conditions. The liionpack framework~\cite{tranter2022liionpack} builds on top of PyBaMM, allowing users to scale up their simulations from a single battery cell to a battery pack of multiple lithium-ion batteries. Although both of these libraries allow for state-of-the-art battery modeling, they do not provide an interface to allow coupling with other simulators for a power system simulation. \newline

\section{Integrating Battery Models into Discrete-Event Co-Simulations}
\label{sec:battery_simulator_integration}

One of the main goals of this work is to describe a common interface for energy storage models to be used inside a microgrid co-simulation. This interface should be adaptable enough so that the complexity range of these models is not too limited, but has to be simple enough to allow a clear co-simulation architecture that still separates power consumption and power generation from the energy storage process, and still provides visibility and control over the energy system.

\subsection{Vessim Architecture}
Vessim~\cite{Wiesner2024_vessim} allows users to define individual microgrids to represent the power systems of computing infrastructure like data centers, utilizing the Mosaik co-simulation framework~\cite{Mosaik_2019}. Each microgrid can consist of multiple simulator responsible for modeling generators, consumers, and energy storage as well as user-defined controllers, which can provide applications visibility and control over the power system through software-in-the-loop simulation~\cite{wiesner2023sil}.

The Mosaik Scenario API defines the simulation architecture of a single microgrid, and the individual simulation units implement the high-level Python API of Mosaik. A Vessim microgrid consists of four distinct subsystems that are stepped in a specific order: Actors, Grid, Controllers, and Storage, whereas the Mosaik Scheduler handles data passing and synchronization.

\begin{figure}
    \centering
    \includegraphics[width=1\linewidth]{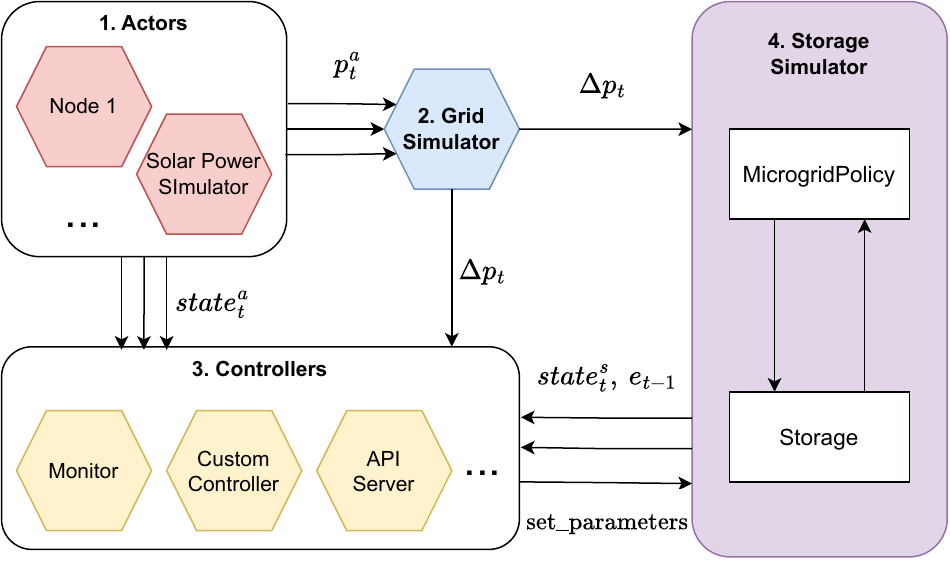}
    \caption{Co-simulation architecture for step-wise execution of battery models, based on Vessim~\cite{Wiesner2024_vessim}.}
    \vspace{-2mm}
    \label{fig:vessim_extension}
\end{figure}

We achieve our described goals by refining the existing co-simulation architecture to include an energy storage interface, energy management policies, and a connection between the controllers and the storage simulator, enabling control of the simulated battery system during runtime. The resulting architecture of a Vessim microgrid is shown in Figure~\ref{fig:vessim_extension}.

\subsection{Refined Storage Simulator}
We add a \textit{Storage} and a \textit{MicrogridPolicy} object to the Vessim storage simulator, which is the last simulator to step, modeling the progression of time between time-steps.

\parag{Battery interface}
The \textit{Storage} standardizes the used battery model, allowing for charging and discharging operations, and containing the monitoring and control functionality of a Battery Management System (BMS). The primary function of the interface is to handle the energy flow into and from the battery, which is facilitated through an update method, receiving the power to be (dis-)charged $p_t^s$ as well as the (dis-)charging duration $d_t^s$.

\parag{Energy management policies}
In the context of microgrid co-simulation, energy management is a critical part, as there have to be ways to define the microgrid's response to specific power deltas, which refer to the variations in power supply and demand. For this purpose, the approach allows for a user-defined \textit{MicrogridPolicy}, which provides a structured way to deal with these power fluctuations, with the management including interactions with the public grid and present energy storage systems as well as curtailment by user-defined rules.

\parag{Control flow}
At every time-step $t$, the storage simulation unit receives $\Delta p_t$ through the Mosaik Scheduler, marking the power delta of the energy producers and consumers, as obtained by the grid simulation unit utilizing a power flow analysis. It then initiates the following control flow:
\begin{enumerate}
    \item The \textit{MicrogridPolicy} receives the power delta $\Delta p_t$ and the time-step duration $d_t$, determining how much of the imbalance should be handled by (dis-)charging the \textit{Storage}.
    \item It instructs the \textit{Storage} to (dis-)charge at power $p_t^s$ for a duration $d_t^s$, which should typically equal $d_t$, but may be shorter in special cases.
    \item The \textit{Storage} updates its internal state accordingly, ensuring battery limits are respected via the integrated BMS.
    \item The \textit{Storage} returns the actual energy stored or discharged, $e_t^s$. If, for example, a charge request of $p_t^s = 20W$ for $d_t^s = 10s$ is limited by the BMS to $15W$, the returned energy is $e_t^s = 15W \cdot 10s = 150Ws$.
    \item The \textit{MicrogridPolicy} computes the net energy exchange with the grid as $e_t = \Delta p_t \cdot d_t - e_t^s$. Assuming no curtailment and continuous grid access, this represents the energy surplus (if $e_t > 0$) or deficit (if $e_t < 0$) since the last time-step.
    \item Finally, $e_t$ is passed to the controllers via the Mosaik Scheduler for the next simulation time-step, indicating whether energy was fed into or drawn from the grid.
\end{enumerate}

\parag{Design choices}
Due to the underlying Vessim architecture, which conducts power flow analysis, the main input to the battery models is power instead of current, which would be the common input to complex models like SPMs or DFN. Additionally, a management system is modeled as part of the battery, similarly to what is described by Kazhamiaka et al.~\cite{kazhamiaka2018pimodel} in their paper discussing the PI-model. The reason behind that is to not allow simulated batteries to be used in ways that would be prevented by a BMS like over- or undercharging, and to allow for monitoring of the battery's internal state. The \textit{MicrogridPolicy} separates itself from the BMS of the energy storage in the way that it can also handle energy flow from or to the public grid, and can be, for instance, used to model the operating mode (grid-connected or islanded) of the microgrid. Because the \textit{MicrogridPolicy} only has access to the abstract energy storage interface, it is also more general, as it does not depend on the battery model in comparison to the BMS.

\subsection{Providing Visibility and Control}
One of the main motivations behind Vessim \cite{Wiesner2024_vessim} was to provide applications visibility and control over the simulated microgrids, which we can achieve the following way:

\parag{Visibility}
The State-of-Charge (SoC) is the key metric when it comes to determining the energy level of a battery, and therefore, the BMS has to give an estimation that can be used for energy management in the microgrid. For other important internal parameters that depend on the type of battery, the \textit{Storage} can also define a custom state $state^{s}$ that can be used for providing visibility over the battery system. This state is also transmitted to the controllers via the Mosaik scheduler. That way, the architecture provides the controllers, and thereby also carbon-aware applications, visibility over the power system, which is one of the key ideas behind the Vessim framework.

\parag{Control}
In terms of the energy management and the energy storage, control means that the controllers can tweak the configuration of the energy storage system during simulation runtime, because a paper by Wiesner et al.~\cite{wiesner2023sil} already describes how applications can interact with the controllers via an API through Software-in-the-loop simulation. Controllers can therefore send a collection of key-value pairs to the storage simulation unit via the Mosaik Scheduler, where they are handled so that relevant functions for the updating of parameters are called on the \textit{MicrogridPolicy} or \textit{Storage} objects. Thus, the responsibility for defining the parameters that can be updated during runtime lies with the developer of the models and policies.

\section{Battery Models}
\label{sec:models}

\begin{figure}[t]
    \centering
    \includegraphics[width=\linewidth]{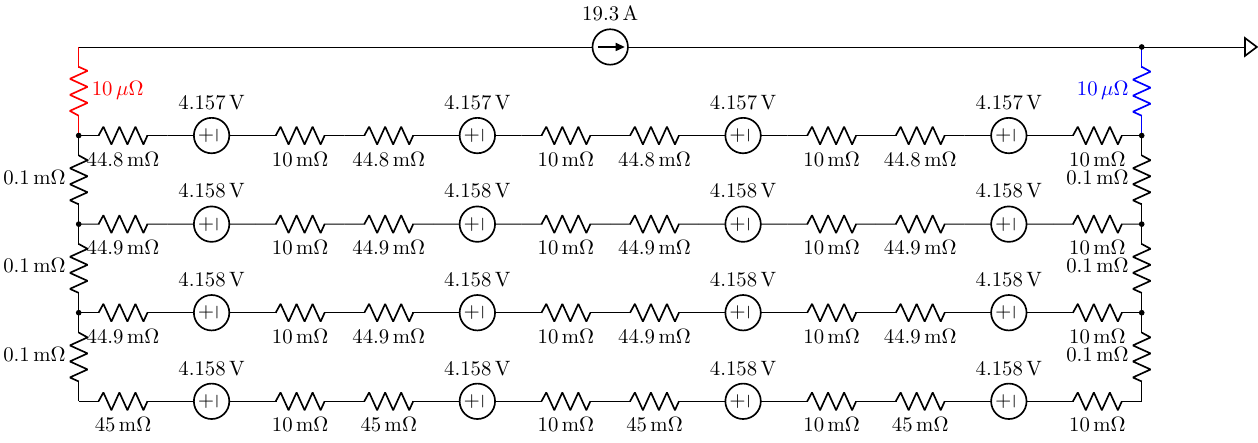}
    \caption{Visualization of a 4S4P battery pack, as obtained utilizing functions from the Liionpack~\cite{tranter2022liionpack} and Lcapy libraries~\cite{lcapy} early during a discharging experiment. Voltage sources and a resistor to their left, indicating an internal resistance, represent the individual battery cells.}
    \label{fig:pack_circuit}
\end{figure}

We compared four different mathematical models of rechargeable lithium-ion battery packs in the extended Vessim framework by parameterizing them to represent a group of INR21700 M50 cells. Figure~\ref{fig:pack_circuit} depicts an exemplary pack of 16 cells, where four series groups of four cells are connected in parallel (4S4P).

There are four different mathematical models described in total that increase in complexity. The \textit{SimpleBattery} is the simplest lossless model, the \textit{CLCBattery} adds power constraints and inefficiencies, the \textit{PybammBattery} considers electrochemistry of individual cells as provided in the PyBaMM framework~\cite{Sulzer_PyBaMM_2021}, and the LiionBatteryPack extends this to include losses due to the battery pack's circuit with methods of the liionpack framework~\cite{tranter2022liionpack}. 

The simple models are parameterized using PyBaMM simulations next to the cells' product specifications to ensure comparability between the different models, and to determine whether the simple linear models are able to capture the physics-based models' behavior.

\subsection{SimpleBattery}
\label{simple_battery}
The \textit{SimpleBattery} only has the battery's energy capacity $C$ in $Wh$ and the initial SoC as parameters, which are used for determining the battery's initial energy $b(0)$. When updated, given the power $p_s(t)$ and the duration $d_s(t)$ of that time-step $t$, the battery's new energy is computed using this formula:
\begin{equation}
b(t) = \begin{dcases}
    \min(C,\: b(t-1) + p_s(t) \cdot d_s(t)) & \vert\: p_s(t) \ge 0\\
    \max(0,\: b(t-1) + p_s(t) \cdot d_s(t)) & \vert\: p_s(t) < 0\\
\end{dcases}
\end{equation}

The (dis-)charged energy that is to be returned when implementing the proposed storage interface, is $p_s(t) \cdot d_s(t)$ in the default case, $-b(t-1)$ if the battery is fully discharged, or $C - b(t-1)$ if the battery is fully charged during the time-step. The formula $\text{SoC}(t) = \frac{b(t)}{C}$ determines the battery's state of charge.

 A complete discharge at $0.2C$ of a single cell returned an energy of $C = 18.87Wh$ in a PyBaMM simulation compared to the $18.20Wh$ that is listed in the product specification. This value is just multiplied by the number of cells to get the total energy capacity $C$ of a battery pack.

\subsection{CLCBattery}
\label{clc_battery}
The \textit{CLCBattery} implements the C-L-C model by Kazhamiaka et al.~\cite{kazhamiaka2019tractable}. When updated, the model behaves similar to the \textit{SimpleBattery}, but includes constant inefficiency factors $\eta_c$ and $\eta_d$, so that the updated energy level can be computed like so:
\begin{equation}
    b(t) = \begin{dcases}
    b(t-1) + \eta_c \cdot p_s(t) \cdot d_s(t) & \vert\: p_s(t) \ge 0\\
    b(t-1) + \eta_d \cdot p_s(t) \cdot d_s(t) & \vert\: p_s(t) < 0\\
\end{dcases}
\end{equation}

The \textit{CLCBattery} adds a lot of complexity regarding battery limitations, though. The assumed constant voltage $V$ and linear energy limits $a_1(I(t)) = u_1 \cdot I(t) + v_1$ and $a_2(I(t)) = u_2 \cdot I(t) + v_2$,  result in the following power limitations when charging or discharging at each time-step $t$:
\begin{align}
    \alpha_d \cdot V &\le p_s(t) \le \alpha_c \cdot V \\
    \frac{b(t-1) - v_1}{\frac{u_1}{V} - d_s(t) \cdot \eta_d} &\le p_s(t) \le \frac{b(t-1) - v_2}{\frac{u_2}{V} - d_s(t) \cdot \eta_c}
\end{align}
These constraints can also be scaled linearly by the number of cells when considering battery packs.

Whenever the input power $p_s(t)$ would exceed these (dis-)charging limits, the BMS simply continues the (dis-)charge at the maximum allowed power. The total energy stored or discharged to be returned is therefore given by $e(t) = p_s(t) \cdot d_s(t)$. Similar to the \textit{SimpleBattery}, the SoC can be easily computed as $\text{SoC}(t) = \frac{b(t)}{v_2}$.

The C-L-C model assumes constant Voltage $V(t)$, and, for the experiments, it is set to the nominal voltage of the modeled lithium-ion cell, which is $3.63V$ according to the product specification. The (dis-)charging current limits were defined to be $\alpha_d = -1.5C$ and $\alpha_c = 0.7C$ at $25^{\circ}C$ in the same way.

PyBaMM simulations, determined the constant inefficiency factors $\eta_c$ and $\eta_d$ based the internal battery resistance $R_i$ using the formula described by Kazhamiaka et al.~\cite{kazhamiaka2019tractable}:
\begin{equation}
    \eta_c(I, V) = 1 - \frac{I \cdot R_i}{V}
\end{equation}
The approximations for the inefficiencies are the values of these functions at the standard discharge rate of $0.2C$, and the standard charge rate of $0.3C$ at the nominal voltage of $3.63V$. The resulting inefficiencies for the cell described by the \textit{CLCBattery} are thereby fixed at $\eta_d = 1.014$, and $\eta_c = 0.978$.

Finally, the linear approximations of the upper and lower energy limits were again determined by PyBaMM simulations:
$$a_1(I) = -0.087 \cdot I \text{,}\:\:\:\: a_2(I) = -1.326 \cdot I + 19.14$$
For battery packs, these energy limits can be scaled linearly by the number of individual cells as well.

\subsection{PybammBattery}
\label{pybamm_battery}
The \textit{PybammBattery} uses the PyBaMM framework~\cite{Sulzer_PyBaMM_2021} directly to simulate an individual lithium-ion cell. For the experiments, the Single Particle Model with electrolyte is used together with an external circuit submodel to enable power-based inputs instead of current-based inputs. The single cell model utilizes the parameterization for the Lithium-Ion INR21700 M50 cell as experimentally obtained by Chen et al.~\cite{chen2020pybammParameterization}. To simulate battery packs, the input power $p_s(t)$ of time-step $t$ is divided by the number of cells before the PyBaMM simulation is stepped. 

\begin{figure}[t]
    \centering
    \includegraphics[width=\linewidth]{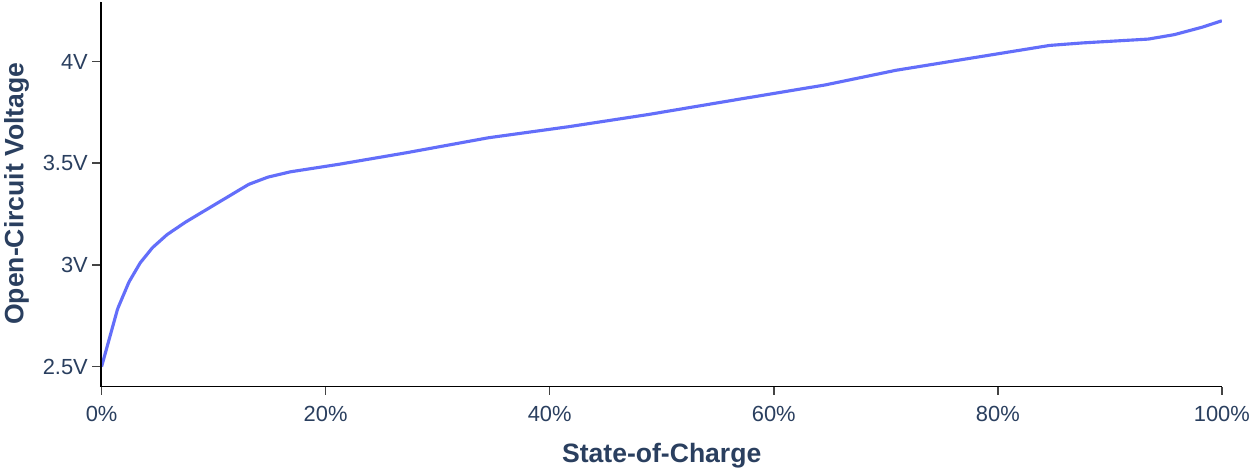}
    \caption{Relation between the battery cell's State-of-Charge and Open-Circuit Voltage as obtained using PyBaMM.}
    \label{fig:soc_estimation}
\end{figure}

The model is equipped with BMS functionality to ensure that battery cells do not exceed their limits, and visibility into the battery’s state is provided. Firstly, the battery's Open-Circuit Voltage (OCV), which is the battery's difference in potential when resting, is used together with an OCV–SoC lookup table, as visualized in Figure~\ref{fig:soc_estimation}, to determine the SoC. This only offers a reasonable estimation of the SoC because the precise OCV is known in the model. In a real-world scenario, this approach is not feasible, as the OCV cannot be measured that accurately, and the relationship between the OCV and the SoC changes with temperature and battery age~\cite{meng2017overview}.

\begin{figure}[t]
    \centering
    \includegraphics[width=\linewidth]{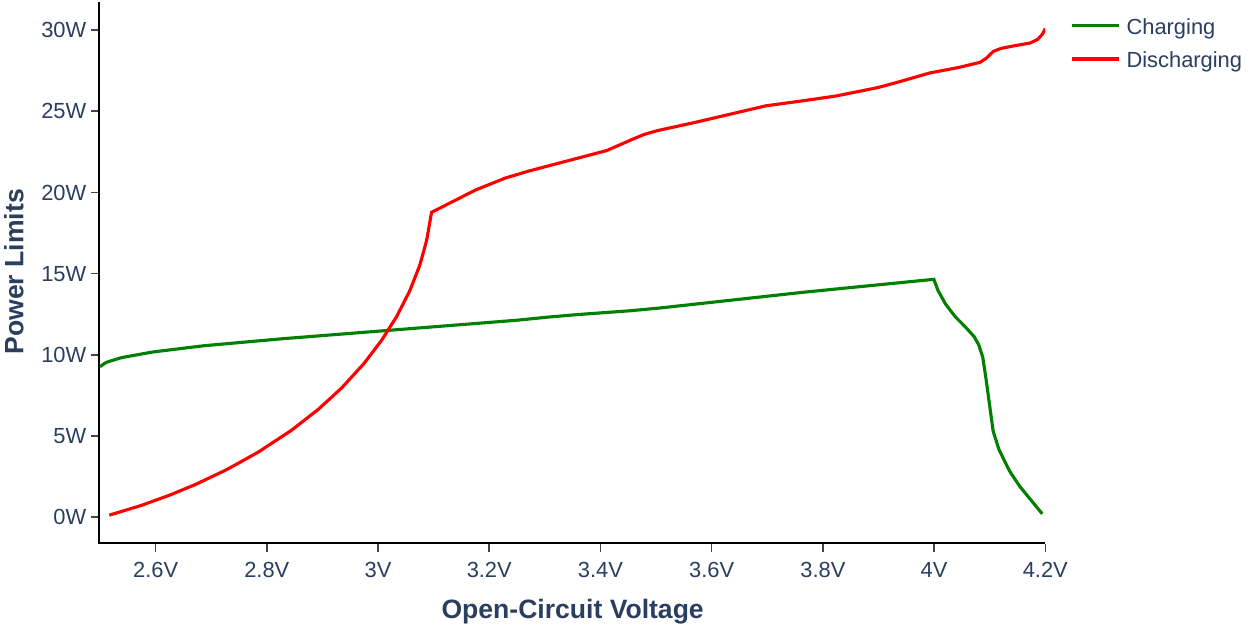}
    \caption{Power limits for charging and discharging based on the cell's OCV. In the more linear parts, the limits are the result of current boundaries, whereas voltage boundaries cause power limitations at a low SoC when discharging and at high SoC when charging.}
    \label{fig:power_limits}
\end{figure}

Secondly, the OCV also determines power limits to ensure that the battery does not exceed allowed currents and voltages when being (dis-)charged with a certain power. In particular, the currents cannot exceed $1.5C$, equivalent to $7.5A$, for discharging, and $0.7C$, equivalent to $3.5A$, for charging, whereas the battery cell's voltage has to remain between $2.5V$ and $4.2V$. Two PyBaMM simulations determine the power limits depending on the OCV, with Figure~\ref{fig:power_limits} showing the results of the experiment. When the input power $p_s(t)$ would exceed the power limits based on the state, the maximum applicable power is used for (dis-)charging and the computation of the stored or discharged energy, which has to be returned.

\subsection{LiionBatteryPack}
\label{liion_battery_pack}
The \textit{LiionBatteryPack} utilizes PyBaMM~\cite{Sulzer_PyBaMM_2021} for simulation of battery cells, as well as methods for computing power-losses and solving power networks by Liionpack~\cite{tranter2022liionpack} to model an entire lithium-ion battery pack, such as the one visualized in Figure~\ref{fig:pack_circuit}. It performs the following series of steps to determine the new battery state at every simulation time-step:
\begin{enumerate}
    \item The internal resistances and voltages are computed for every individual cell and included in the network representation.
    \item The network is solved based on the input power $p_s(t)$, which determines the applied currents for the cells.
    \item The simulation steps the electrochemical system of each cell with their respective current input using PyBaMM.
\end{enumerate}

Again, a BMS wraps the model. It computes the SoC for all the cells in the same way as for the \textit{PybammBattery}, and then averages them in the end to obtain the SoC of the battery pack. The power limits are also determined similar to the PyBaMM model described in Section~\ref{pybamm_battery}, as the model's power limits are computed for the cell with the highest voltage when charging and for the cell with the lowest voltage when discharging, before being scaled by the number of batteries in the battery pack. The determined applied power is once again used to compute the returned (dis-)charged energy $e(t)$ by multiplication with the duration $d_s(t)$.

\section{Experimental Results}
\label{sec:evaluation}

The described models are now used to represent a 16S16P battery pack of the INR21700 M50 cells for a series of experiments. We determine the behavior of the models in simulated scenarios of constant charging and discharging, before presenting the results, when using the models in a simplified data center scenario. Finally, we perform a runtime analysis.

\subsection{Constant Power (Dis-)Charging}
\label{constant_power_(dis-)charging}
\begin{figure}[t]
    \centering
    \includegraphics[width=\linewidth]{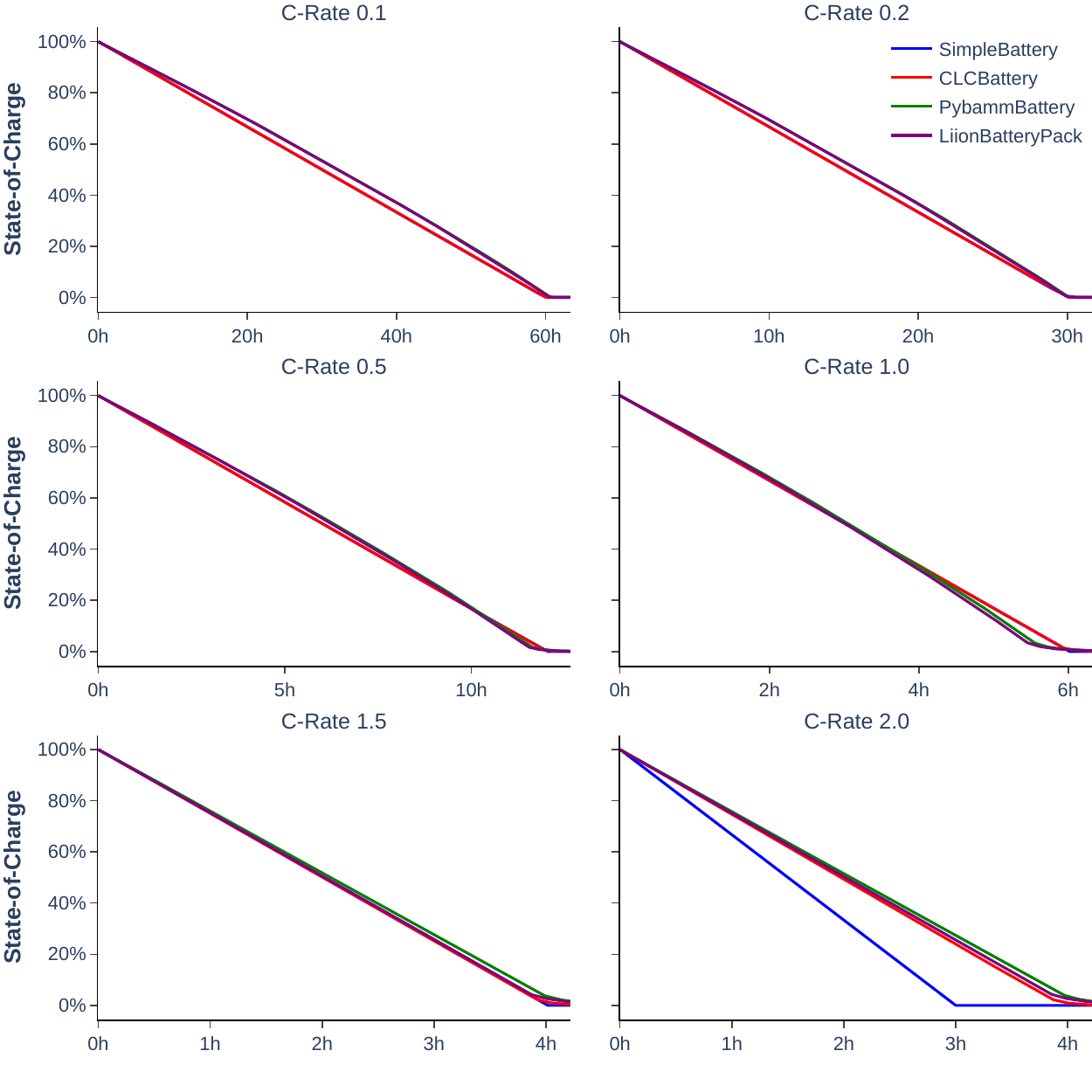}
    \caption{Behavior of the models when discharged at constant power with different rates.}
    \label{fig:discharging_experiment}
\end{figure}

\begin{figure}[t]
    \centering
    \includegraphics[width=\linewidth]{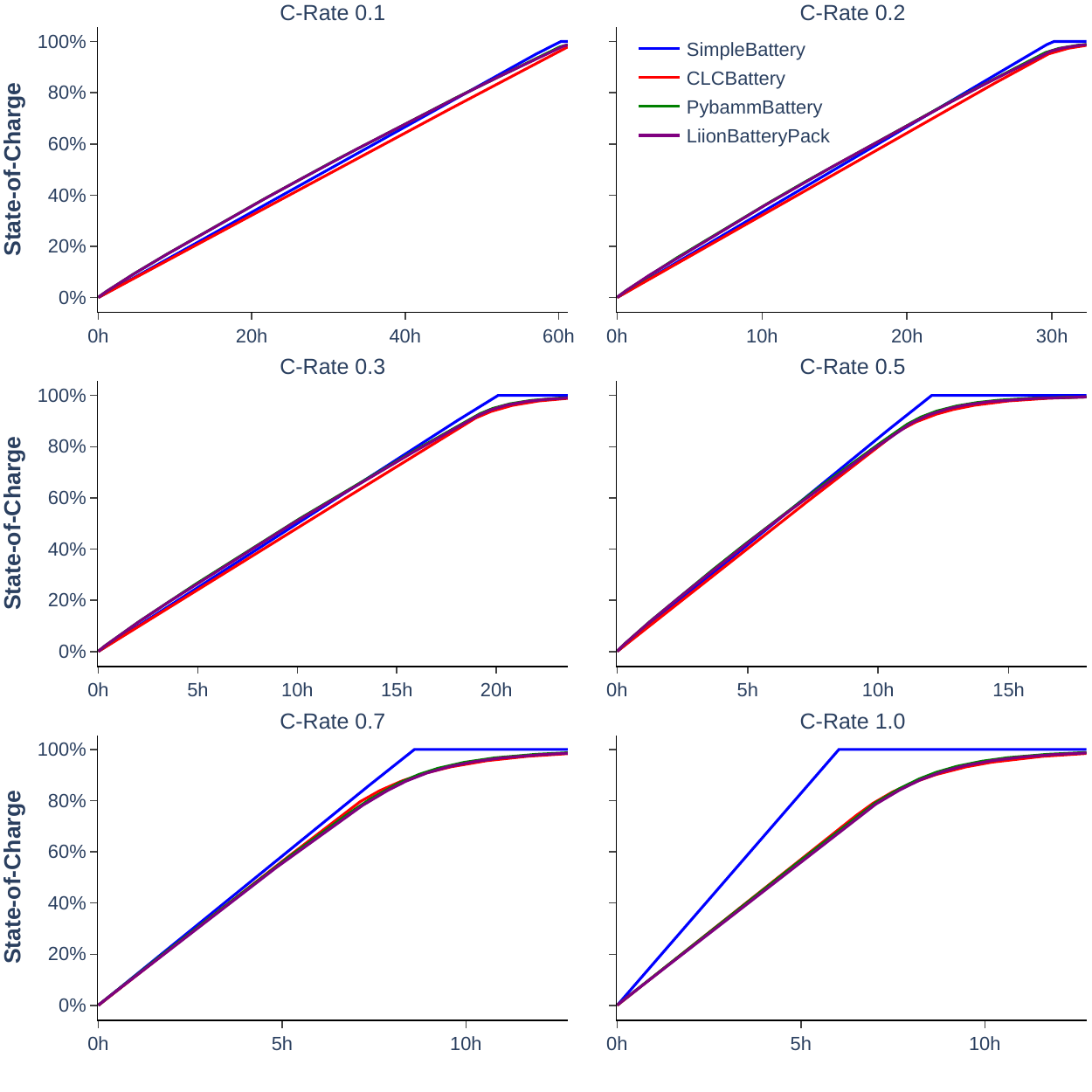}
    \caption{Behavior of the models when charged at constant power with different rates.}
    \label{fig:charging_experiment}
\end{figure}

Figure~\ref{fig:discharging_experiment} and Figure~\ref{fig:charging_experiment} visualize the results of constant power discharging from full to empty state and constant power charging from empty to full state. Because of inaccurate SoC estimations of the complex models, we cannot exactly quantify the SoC differences, and instead focus on the overall trends.

\parag{Discharging}
When comparing the \textit{SimpleBattery} to the \textit{CLCBattery} in the discharging scenarios, it is noticeable that they produce almost the same traces, as both of their capacities were obtained using the drawn energy received by PyBaMM discharging experiments. Because the lower energy limit of the \textit{CLCBattery} is not rising steeply, a real difference at low SoC values can only be seen at high discharge rates. As soon as these rates exceed $1.5C$, the \textit{CLCBattery} limits the output power, resulting in the faster discharging of the \textit{SimpleBattery}.

The \textit{CLCBattery} achieves fairly accurate results relative to the \textit{PybammBattery} and the \textit{LiionBatteryPack} over all discharging rates. It is, however, visible that it empties slightly faster at low rates because the constant $\eta_d$ factor overestimates the actual inefficiencies in these cases. The higher the applied power, the more the inefficiency is underestimated, which results in the slower discharges, when no current limitations are applicable. At $1.5C$, when limitations are in place, the \textit{CLCBattery} once again empties faster than the complex models due to the voltage of their cells being lower here next to the nominal voltage the \textit{CLCBattery} uses, resulting in smaller power limits. 

The \textit{PybammBattery} reflects the \textit{LiionBatteryPack} almost perfectly, as only at higher discharging rates the power losses of the circuit, which are determined in the later model, have a high enough impact to be visible in the overall trend. Through this experiment, it can also be observed that in both of the models, the SoC is likely to be overestimated ever so slightly around half-full state. This imprecision in the determination of the SoC is also one of the major disadvantages of these models.

\parag{Charging}
The differences between the \textit{SimpleBattery} and the \textit{CLCBattery} become more obvious for the charging experiments. Even though they initially charge at the same speed at low to medium charging rates, due to the quickly falling upper energy limits of the \textit{CLCBattery}, the \textit{SimpleBattery} reaches its full state much sooner, as no power limitations are considered here. At charging rates exceeding the set $0.7C$, the BMS of the \textit{CLCBattery} limits the current as well, resulting in an even slower discharge.

The \textit{CLCBattery} seems to reflect the charging process of the \textit{PybammBattery} and the \textit{LiionBatteryPack} quite well again over all charging rates. Similar to the discharging experiment, the \textit{CLCBattery} overestimates the inefficiency at low charging rates, and underestimates the inefficiency at higher charging rates. The underestimation is not as apparent in the figure due to aforementioned inaccuracies of the SoC estimation for the complex models.

Because the current limits, and the resulting power limits, are even lower for charging than for discharging, the differences between the \textit{PybammBattery} and the \textit{Liionpack} are even smaller in this experiment, as the power losses induced by the pack's circuit are almost negligible.

\subsection{Data Center Scenario}
While utilizing the enhanced co-simulation capabilities of Vessim~\cite{Wiesner2024_vessim}, this section conducts an experiment over two days based on a more realistic system configuration that is supposed to represent the power system of a simplified data center. The goal is to test the models in a more variable environment under real operating conditions, similar to an experiment by Wiesner et al.~\cite{wiesner2023sil}.

\begin{figure}[t]
    \centering
    \includegraphics[width=\linewidth]{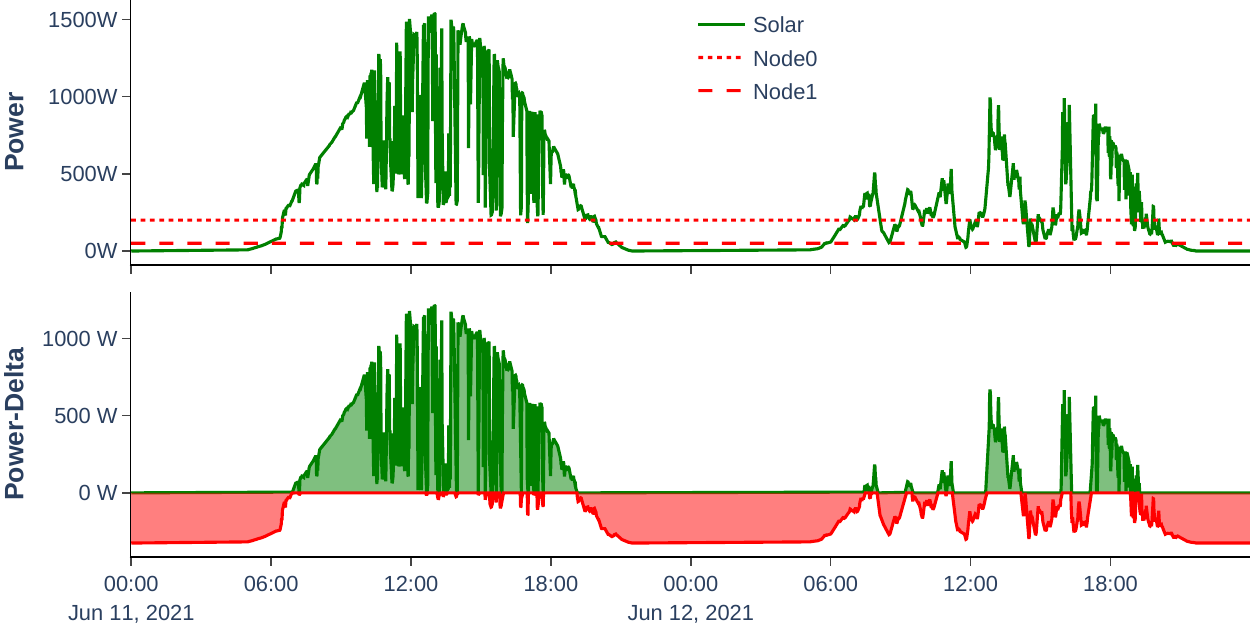}
    \caption{Power production, consumption, and power delta, as used in the experiment.}
    \label{fig:scenario_power}
\end{figure}

\parag{Scenario}
The “data center” contains two nodes, which represent servers, that have a constant power consumption of $200W$ and $50W$ respectively, which are aggregated and multiplied by a power usage effectiveness (PUE) of $1.3$ to obtain the total power consumption of the mocked data center. Additionally, the power production of a solar panel of $10m^2$ size with an efficiency of $15\%$ is simulated, using data of global solar radiation in Berlin from June 2021 \footnote{The data was retrieved from \url{https://wetter.htw-berlin.de}}. Figure~\ref{fig:scenario_power} presents the resulting power traces.

In the scenario, the co-simulated microgrid is allowed to exchange energy with the public grid at all times, and the included \textit{MicrogridPolicy} tries to (dis-)charge as much of the power-delta as possible using the \textit{Storage}, while exchanging the remaining energy delta with the grid. On top, the policy allows a minimum SoC to be set, at which point it stops to further discharge the energy storage, and includes functionality to charge the \textit{Storage} at a specific power rate using the power-delta and grid-energy.

During all the simulations, a custom controller instructs the \textit{MicrogridPolicy} to charge the battery pack at $3200W$ between 11AM and 12AM on the second day due to insufficient solar production. We fix the minimum SoC to $30\%$ at all times to ensure operation of the data center in case of grid outages.

\begin{figure}[t]
    \centering
    \includegraphics[width=\linewidth]{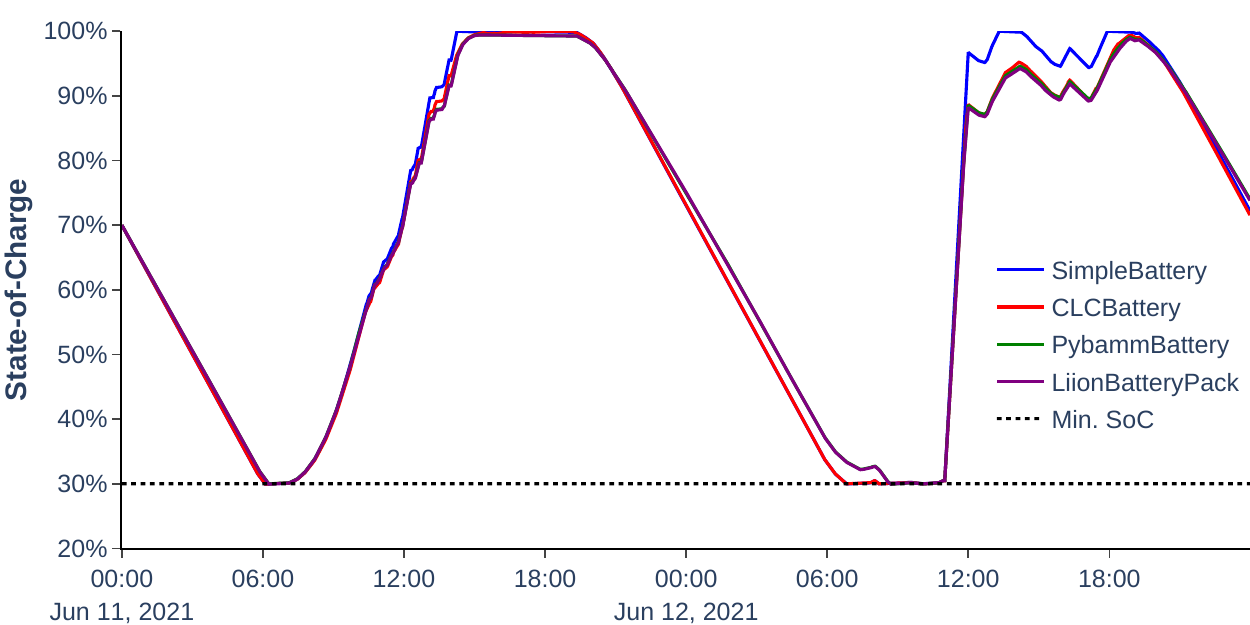}
    \caption{SoC Progression of the models during the experiment.}
    \label{fig:scenario_soc}
\end{figure}

\parag{SoC estimation}
Figure~\ref{fig:scenario_soc} represents the development of the SoC, depending on the model used, and the trends show the discovered characteristics outlined in Section~\ref{constant_power_(dis-)charging}. In general, while the \textit{SimpleBattery} fails to accurately represent the SoC of the more complex models, in particular when charging at a high SoC, the \textit{CLCBattery} captures the physical models quite well throughout the experiment. Again, no real differences in the estimation can be seen between the \textit{PybammBattery} and the \textit{LiionBatteryPack}. Similar to the constant (dis-)charging experiments, an accurate quantification of differences in SoC levels is not possible due to estimation errors for the complex models. 

\begin{figure}[t]
    \centering
    \includegraphics[width=\linewidth]{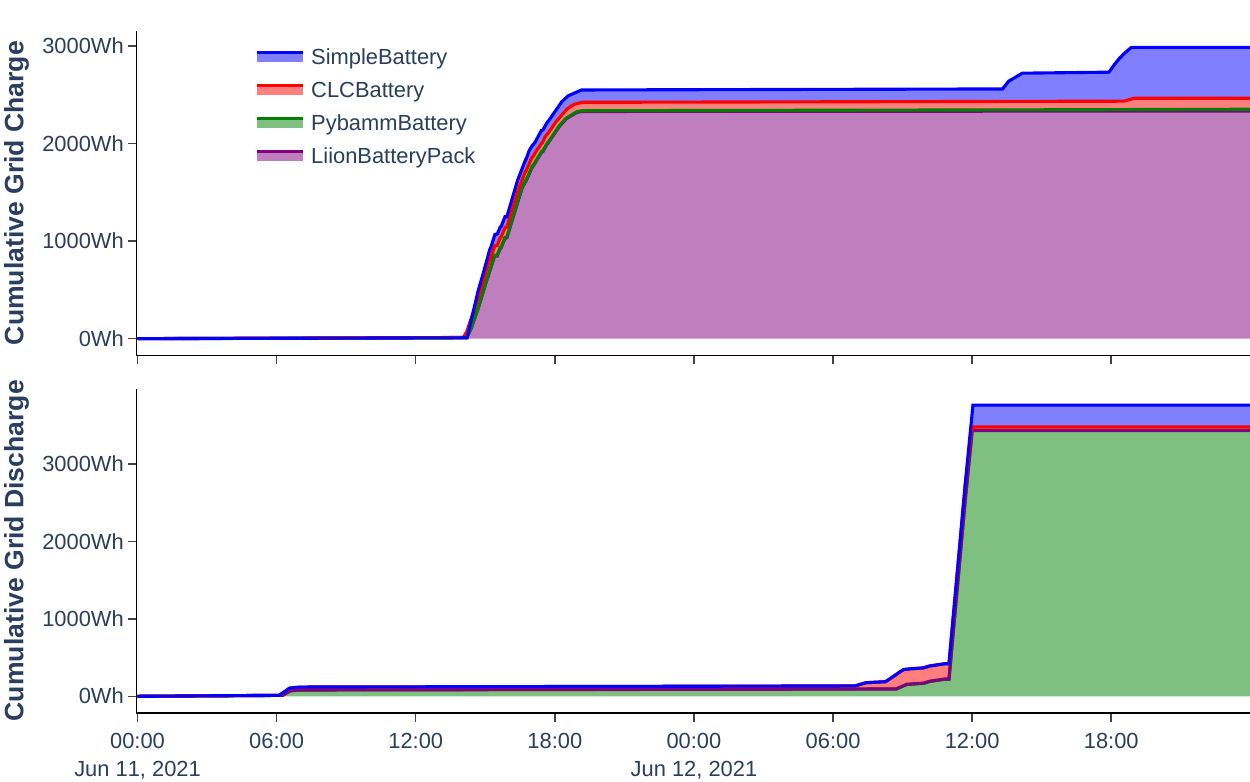}
    \caption{Exchanged power with the grid over the experiment. Grid discharge indicates energy used, while grid charge refers to energy fed to the utility grid.}
    \label{fig:scenario_grid_energy}
\end{figure}

\parag{Grid energy}
When looking at the exchanged energy with the grid (depicted in Figure~\ref{fig:scenario_grid_energy}), we can obtain that over the course of the two days, the \textit{LiionBatteryPack} used a net $1103Wh$ of energy from the grid, with the battery having to charge a total amount of $3435Wh$, and having fed back only around $2332Wh$. This net grid charge is $1.9\%$ higher than for the \textit{PybammBattery} ($1081Wh$), $8.5\%$ higher than for the \textit{CLCBattery} ($1016Wh$), and a total of $41.5\%$ higher than for the \textit{SimpleBattery} ($779Wh$) for this particular scenario.

Because the \textit{SimpleBattery} does not consider any inefficiencies, and is parameterized to best reflect the SoC of the more complex models, it charges and discharges completely using less energy so that more energy exchange with the grid is necessary. This is obviously not the case for the \textit{CLCBattery}, hence the good approximation of the physical models. The remaining inaccuracy here comes from over- and underestimations of inefficiencies, and slightly different power limits, meaning that for a good parameterization of the C-L-C model, the difference in exchanged grid energy is likely to balance relative to the used \textit{PybammBattery} over long experiments. When looking at the \textit{PybammBattery} against the \textit{LiionBatteryPack}, we can determine that the power losses of the pack's circuit do not make a huge difference for short-term experiments.

\subsection{Runtime Analysis}
\label{runtime_analysis}
\begin{figure}[t]
    \centering
    \includegraphics[width=\linewidth]{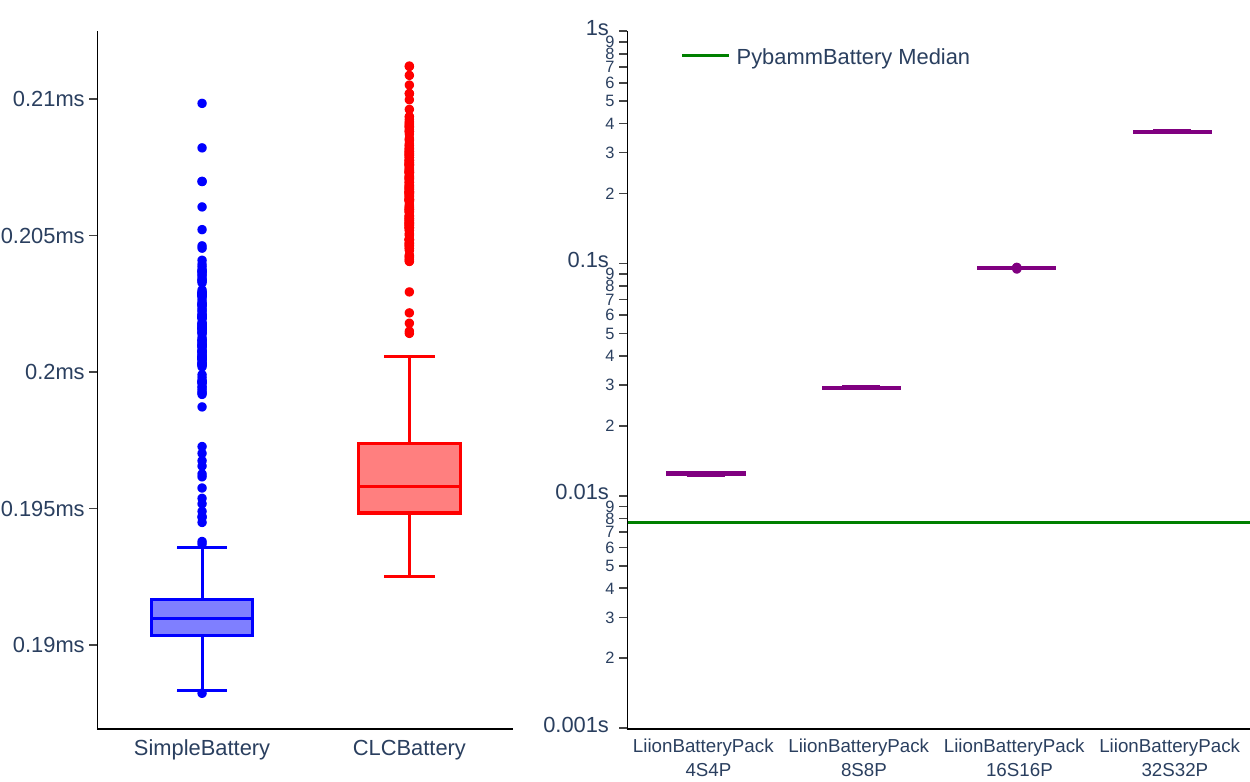}
    \caption{Runtime distributions for execution of a single simulation time-step.}
    \label{fig:runtime_analysis}
\end{figure}

To determine the efficiency and therefore the usefulness of the described models, we conduct a runtime analysis. A virtual node in the cloud \footnote{Machine type: Google Cloud Platform GCE C3} executes the runtime experiments for all models to ensure comparability of the obtained data. Figure~\ref{fig:runtime_analysis} presents the experimental results.

As was to be expected, there is only little difference in performance between the \textit{SimpleBattery} and the \textit{CLCBattery} because both are linear models. The median execution time of a single simulation time-step for the \textit{SimpleBattery} was around $191\mu s$ compared to $196\mu s$ of the \textit{CLCBattery}. For both these models and the \textit{PybammBattery}, the execution time does not change with the size of the battery pack, as they use a single cell modeling approach, where relevant parameters are scaled using the number of cells in the battery pack.

The complexity of the optimization problem for the \textit{LiionBatteryPack} scales linearly with the number of cells, as can be seen in the figure. A $32S32P$ battery pack imitated using the \textit{LiionBatteryPack} model with a total of $1024$ cells can only step two to three times per second on the used node, whereas the model for the $16S16P$ pack of earlier experiments has a median execution time of $0.095s$. That is nearly $500$ times higher than for the \textit{SimpleBattery} and the \textit{CLCBattery}, and more than $12$ times higher than for the \textit{PybammBattery} with $7.65ms$.

\section{Discussion}
\label{sec:discussion}

In this section, we discuss the impact of our results and limitations with our approach.

\parag{Architecture opportunities}
Our experiments show that the extended Vessim co-simulation architecture with the refined storage simulation unit (Section~\ref{sec:battery_simulator_integration}) allows the integration of simple to complex battery models, as shown by the variety of models used for experimentation. According to the results from the data center scenario, the proposed extension can be used to study energy- and battery management inside isolated power systems like microgrids, and the resulting architecture can therefore serve as a co-simulation testbed for carbon-aware approaches in data center scenarios.

\parag{Linear models}
When we compare the models, it becomes obvious that for the modeled lithium-ion cell, the \textit{SimpleBattery} fails to accurately represent the battery's SoC, especially for charging at an almost full state, while also providing inaccurate estimations of exchanged grid energy in a microgrid scenario. This makes it very unsuitable for experiments with the main focus on energy management. Because of the straightforward parameterization, the fast runtime, and the independence of the form of energy storage, it is still useful for early investigation phases of new approaches.
Compared to the \textit{SimpleBattery}, the \textit{CLCBattery} has only minimal performance overhead, while being able to estimate a battery's state, and used grid energy similar to physics-based models. The experiments indicate that the C-L-C model with a decent parameterization has the potential to accurately represent a lithium-ion cell's behavior in short-term experiments. It is therefore applicable for most use-cases utilizing microgrid simulation over a short timeframe, which is the case for a big chunk of experiments in the area of sustainable computing. A representative parameterization for the utilized battery cell can be easily found using product specifications or other experimental data, and even tools for battery simulation like PyBaMM~\cite{chen2020pybammParameterization}.

\parag{Electrochemical models}
Even though SPMs are considered to be the simplest electrochemical models~\cite{marquis2019spm, ramadesigan2012modeling}, solving them still comes with a big computational cost, although the single-cell modeling approach provides good scalability for large battery packs. A decent parameter set can, however, only be obtained through extensive real-world experiments, with frameworks like PyBaMM fortunately providing them for a few cell types. On top, there are significant difficulties in obtaining a useful estimation of the cell's SoC, and models are very dependent on the type of battery. These types of models are therefore not useful for most short-term scenarios involving co-simulated microgrids, as these representations of power systems possibly bring additional inaccuracies, making the utilization of physics-based models not justifiable. There exists, however, the opportunity to investigate the behavior of such electrochemical models in long-term experiments, when aspects like battery degradation in data center scenarios are of interest, because models such as SPMs can be extended with coupled degradation mechanisms as investigated by O'Kane et al.~\cite{o2022lithium}.

For the battery packs investigated in this work, there does not seem to be a payoff in modeling a whole battery pack instead of a scaled single cell, as the discovered effects of power losses were fairly minimal. For larger battery packs, the execution times become too substantial, even up to the point where it would be critical for real-time applications, but the stepping of the different electrochemical battery cells has parallelization opportunities to solve this runtime issue. All in all, the results of the experiments using the \textit{LiionBatteryPack} have the most uncertainty to them, yet the collected data questions the usefulness of modeling a battery pack in a way that considers electrochemical cells and the pack's circuit.

\parag{Limitations}
Although electrochemical battery models, such as SPMs, effectively describe battery behavior, they inherently face uncertainty due to model assumptions and parameter estimation errors~\cite{Escalante_2021_DFN_Uncertainty, Aitio_2020_SPMe_bayesian_parameterization_uncertainty}. As a result, their predictive accuracy under real-world conditions remains limited.

Furthermore, co-simulation testbeds for microgrids introduce additional challenges. They often rely on simplified assumptions about battery pack characteristics, which may not fully capture the complexity of large-scale deployments, such as those in data center microgrids. The lack of high-fidelity data on aging effects, thermal behavior, and variability across battery units further increases uncertainty. Consequently, while these models provide valuable insights, their applicability to large-scale, real-world microgrids requires careful validation against empirical data.

\section{Outlook}
\label{sec:outlook}

As a next step, we aim to extend our analysis by incorporating recent battery degradation models~\cite{BLASTLite, o2022lithium}, which estimate battery lifetime based on factors like temperature, state of charge, depth-of-discharge, and (dis)charge rates. 
Integrating these models into data center co-simulations can significantly improve control strategies and long-term investment planning.
Additionally, we plan to leverage our battery simulator integration in systems research, particularly for developing and testing energy-aware federated learning systems with battery-powered clients~\cite{arouj2022fl_batterypowered, wiesner2024fedzero}. Beyond lithium-ion storage, we aim to extend our research to alternative energy storage solutions, such as hydrogen-based storage systems~\cite{Lazaar_HydrogenDataCenter_2021, NREL_HydrogenDataCenter_2019, SEPIA_DATAZERO_2019}.

Given the limitations of our current evaluation, future work should focus on establishing reliable benchmarking methods for both high-level battery models and co-simulated energy systems, ensuring their accuracy and applicability to real-world data center environments.

\bibliographystyle{ACM-Reference-Format}
\bibliography{bibliography}
\end{document}